\documentclass[aps,floatfix,twocolumn,footinbib,superscriptaddress]{revtex4}
\usepackage{epsfig}
\usepackage{amsmath,amssymb}
\usepackage{bm}
\begin{document}

\title{Proposal for a ferromagnetic multiwell spin oscillator}

\author{Christian Ertler\footnote{email:christian.ertler@uni-graz.at}}
\affiliation{Institute of Theoretical Physics, Karl-Franzens University Graz,
Universit\"atsplatz 5, 8010 Graz, Austria}
\affiliation{Institute for Theoretical Physics, University of
Regensburg, Universit\"atsstrasse 31, D-93040 Regensburg, Germany}
\author{Walter P\"otz}
\affiliation{Institute of Theoretical Physics, Karl-Franzens University Graz,
Universit\"atsplatz 5, 8010 Graz, Austria}
\author{Jaroslav Fabian}
\affiliation{Institute for Theoretical Physics, University of
Regensburg, Universit\"atsstrasse 31, D-93040 Regensburg, Germany}

\begin{abstract}

The highly nonlinear coupling of transport and magnetic properties
in a multiwell heterostructure, which comprises ferromagnetic
quantum wells made of diluted magnetic semiconductors, is
theoretically investigated. The interplay of resonant tunneling and
carrier-mediated ferromagnetism in the magnetic wells induces very
robust, self-sustained current and magnetization oscillations. Over
a large window of steady bias voltages the spin polarization of the
collector current is oscillating between positive and negative
values, realizing a spin oscillator device.

\end{abstract}

\maketitle

Magnetic resonant tunneling structures are prominent spintronic
devices \cite{Fabian2007:APS}, which are proposed for spin valves,
spin filtering
\cite{Likovich2009:PRB,Slobodskyy2003:PRL,Petukhov2002:PRL,Ohya2010:PRL,
Ertler2006a:APL}, or for realizing digital magnetoresistance
\cite{Ertler2007a:PRB}. In nonmagnetic multiwell heterostructures
interesting dynamic effects such as the formation of electric field
domains and the motion of charge dipoles through the structure can
be observed \cite{Bonilla2005:RPP, Bonilla2006:PRB}. Recently, it
has been predicted that in structures with paramagnetic wells these
phenomena can be controlled by an external magnetic field
\cite{Sanchez2001:PRB,Escobedo2009:NJP, Escobedo2009:PRB}. The
paramagnetic systems can behave as a spin oscillator device, in
which the moving charge dipoles generate spin-polarized current
oscillations in the MHz range \cite{Bonilla2007:APL, Bejar2003:PRB,
Escobedo2009:PRB}. In ferromagnetic multiwell structures made of
dilute magnetic semiconductors (DMS) in addition also the magnetic
order will become dynamic \cite{Ertler2008:PRL}. This additional
degree of freedom enriches the dynamic complexity, which 
up to now is still largely unexplored. DMSs
\cite{Ohno1998:S,Dietl:2007,Jungwirth2006:RMP} are made magnetic by
doping them with transition metal elements, e.g., by
incorporating Mn in a GaAs crystal host. In several experiments
ferromagnetism has been generated in the bulk by tailoring the
actual particle density electrically or optically \cite{Ohno2000:N,
Boukari2002:PRL}. In 2d-confined systems made of DMSs the magnetic
order depends strongly on the local spin density, which is
influenced by the tunneling current
\cite{Dietl1997:PRB,Jungwirth1999:PRB, Lee2002:SST}. This interplay
leads to a novel mechanism of generating robust self-sustained
current oscillations.

In this article a detailed study of the dynamics in a ferromagnetic
four-well system, as a prototype of the expected multiwell functionality, is provided. 
The carrier dynamics is described by
a self-consistent sequential tunneling model which includes
momentum and spin relaxation inside the wells. The feedback effects
of both the carriers Coulomb interaction and the magnetic exchange
coupling with the magnetic ions are described within a mean-field
picture. We give an qualitative explanation for the occurrence of
spin-polarized current oscillations, which realizes the
functionality of a spin oscillator device.

%
%

\begin{figure}
\centerline{\psfig{file=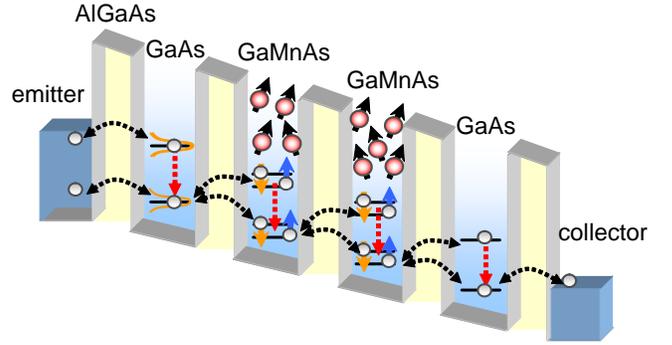,width=1\linewidth}}
\caption{(Color online) Schematic scheme of the band profile of the
ferromagnetic four-well structure. The exchange interaction of the
magnetic ions in the second and third well is mediated by the
carriers tunneling in and out of the wells.} \label{fig:scheme}
\end{figure}

The band profile of a four-well heterostructure with two
ferromagnetic quantum wells made of a DMS, e.g., of GaMnAs, is
sketched in Fig.~\ref{fig:scheme}. In the ferromagnetic wells
the ferromagnetic order of the magnetic ions is mediated by the itinerant carriers.
Describing the mutual exchange interaction within a mean-field picture
allows to derive an analytic expression for the exchange splitting $\Delta^0_i$
of the subbands in the ferromagnetic wells 
\cite{Dietl1997:PRB,Jungwirth1999:PRB, Lee2000:PRB, Fabian2007:APS}
\begin{eqnarray}
\Delta_i^0 &=& J_\mathrm{pd}\int\mathrm{d}z\:
n_\mathrm{imp}(z)\left|\psi_0(z)\right|^2\nonumber\\
\label{eq:exchange}&& \times S B_S\left[\frac{S J_\mathrm{pd} s
(n^\uparrow_i- n^\downarrow_i) \left|\psi_0(z)\right|^2}{k_B T}\right],
\end{eqnarray}
where $J_\mathrm{pd}$  is the coupling strength between the
impurity spin and the carrier spin density (in case of GaMnAs p-like
holes couple to the d-like impurity electrons), $z$ is the
longitudinal (growth) direction of the structure,
$n_\mathrm{imp}(z)$ is the impurity density profile of magnetically active ions,  $\psi_0(z)$
labels the well ground wave function, and $s=1/2$ is the particles spin. The Brillouin function
of order $S$ is denoted by $B_S$, where $S $ is the impurity spin,
which for Mn equals 5/2.
We assume a homogenous impurity distribution making the
spin polarization $\xi_i = s(n^\uparrow_i- n^\downarrow_i)$ the determining factor
for the exchange splitting $\Delta^0_i$.

The well spin polarization can be rapidly changed by in- and out-tunneling particles.
However, the magnetic impurities need some time to respond until the corresponding mean
field value $\Delta^0$ is established. In the case of GaMnAs,
experimental studies of the magnetization dynamics revealed typical
response times $\tau_\Delta$ of about 100 ps \cite{Wang2007:PRL}. We model the
magnetization evolution in the ferromagnetic wells $i=(2,3)$ within the relaxation time approximation,
$\mathrm{d}\Delta_i/\mathrm{d}t = -(\Delta_i-\Delta^0_i)/\tau_\Delta$.

The transport through weakly coupled multiwell structures is well
described by a sequential tunneling model, which is presented in
detail in Refs.~\cite{Bonilla2006:PRB, Escobedo2009:PRB}. The
transport model includes the following assumptions: (i) The wells are
weakly coupled and tunneling can be described by an transfer Hamiltonian
formalism. (ii) The particles in each well always stay in a
quasiequilibrium state due to elastic and inelastic scattering,
which is assumed to constitute the shortest time scale in the
problem ($\tau_\mathrm{scatt} \approx 1$ ps). (iii) Particles in the
first excited subband relax rapidly to the ground level by
 inelastic scattering processes, e.g., due to emission of LO-phonons.
This happens within a few ps, which is much smaller than the other relevant time scales
occurring in the problem. 
(iv) The spectral function of the subbands is a
Lorentzian with the broadening given by $\gamma
=\hbar/2 \tau_\mathrm{scatt}$. (v) During tunneling processes the
parallel momentum and spin are conserved, i.e., effects of interface
roughness are not taken into account. (vi) Spin flipping processes
inside the well can be described by a single spin relaxation time
$\tau_s$. Based on these approximations rate equations for the
spin-dependent particle densities $n^\sigma_i , (\sigma =
\uparrow,\downarrow =\pm 1/2)$ in the $i$th well can be found
\begin{equation}\label{eq:rate}
\frac{\mathrm{d}n^\sigma_i}{\mathrm{d} t} = J^\sigma_{i-1\rightarrow i}-J^\sigma_{i\rightarrow i+1}
-\frac{n^\sigma_i-n^{0,\sigma}_i}{\tau_s}.
\end{equation}
where the tunneling currents $J_{i\rightarrow i+1}$ are derived
microscopically within the Kubo-formalism \cite{Xu2007:PRB,
Bonilla2006:PRB}. The quasiequilibrium particle spin densities
$n^{0,\sigma}_i$ are calculated from the two conditions
$n^\uparrow_i+n^\downarrow_i = n^{0,\uparrow }_i+n^{0,\downarrow
}_i$ and $n^{0,\uparrow }_i-n^{0,\downarrow }_i = \xi^0_i/s$, with
$\xi^0_i$ resulting from inverting Eq.~\ref{eq:exchange} for the
actual $\Delta_i$.

Finally, the electron-electron interaction is taken into account within a discretized Poisson equation, yielding
\begin{equation}
 F_i-F_{i-1} = \frac{e}{\varepsilon}\left(\sum_\sigma n_i^\sigma - N_{\mathrm {back}}\right)
\end{equation}
with $F_i$ denoting the electric field at the $i$th barrier, 
$e$ is the elementary charge, $\varepsilon$ denotes the
dielectric constant, and $N_\mathrm{back}$ labels the background charge.

The tunneling currents are strongly sensitive to the relative
alignment of adjacent subbands and since the quantum well levels
$E^\sigma_{i,j} = E_j+\phi_i-\sigma\Delta_i$ explicitly depend on
the magnetic exchange splitting and the electrostatic potential
$\phi_i$ all equations are coupled highly nonlinearly.

For the numerical simulations we used generic parameters
representing GaAs and GaMnAs quantum wells, respectively, $m = 0.5\:m_0$,
$\varepsilon_r = 12.9$, $V_\mathrm{bar} = 300$ meV, $d = 20$ \AA, $w
= 20$ \AA, $E_0 = 80$ meV, $ E_1 = 188$ meV, $\mu_e = 100$ meV,
$n_\mathrm{imp} = 0.5\times10^{20} $cm$^{-3}$, $J_{\mathrm{pd}} =
0.15$ eV nm$^3$ \cite{Lee2000:PRB}, $\gamma = 2$ meV, and $\tau_s =
0.1$ ns, where $m$ is the effective mass with $m_0$ denoting the
free electron mass, and $\varepsilon_r$ is the relative
permittivity, $V_\mathrm{bar}$ is the barrier height, $d$ and $w$
are the barrier and quantum well widths, $E_0$ and $E_1$ are the
energies of the two lowest subbands, and  $\mu_e$ is the emitter
Fermi energy. The background charge $N_\mathrm{back} =
0.1\:n_\mathrm{imp}w = 10^{12}$ cm$^{-2}$ is considered to be only of about 10\% of the
nominal Mn doping density \cite{DasSarma2003:PRB} and the lattice
temperature is $T = 4.2$~K. The boundary currents
$J_{0\rightarrow 1}$ and $J_{N\rightarrow N+1}$ with $N$ denoting
the total number of quantum wells are determined by the tunneling
from or into a three-dimensional Fermi sea at the emitter and
collector side, respectively. For these currents analytic
expressions can be derived \cite{Xu2007:PRB,
Bonilla2005:RPP}.

%
%
\begin{figure}
\centerline{\psfig{file=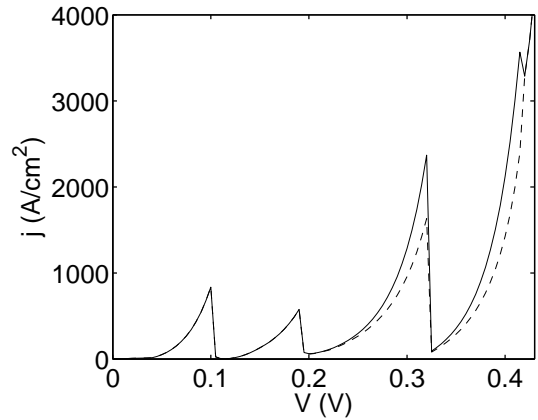,width=0.9\linewidth}} \caption{
Current-Voltage (IV)-characteristics of the four well
structure. Self-sustained current oscillations appear in the 
voltage range of $V = 0.22$ V to $V = 0.42$ V, in which both the maximum and minimum values of
the oscillations are indicated by the solid and dashed lines,
respectively.} \label{fig:IV}
\end{figure}

\begin{figure}
\centerline{\psfig{file=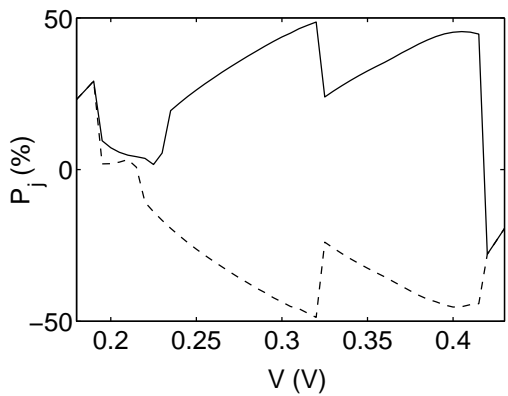,width=0.9\linewidth}} \caption{Spin polarization of the collector current versus applied bias $V$ in the
voltage range in which oscillations are occurring. The maximum and minimum values of the
oscillations are
indicated by the solid and dashed lines, respectively.  } \label{fig:Pj}
\end{figure}

\begin{figure}
\centerline{\psfig{file=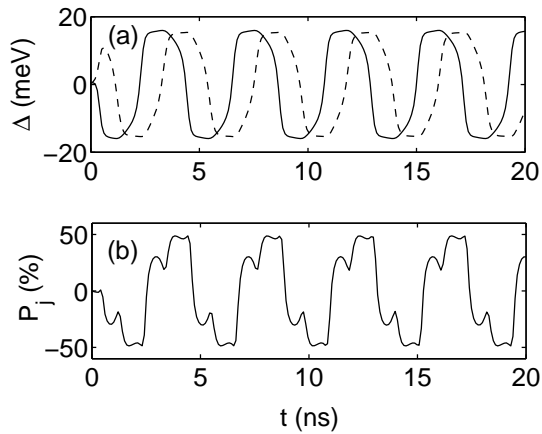,width=0.9\linewidth}}\caption{(a) Exchange splitting $\Delta_i$ in the second (solid line) and
third (dashed line) well versus time $t$ at $V = 0.32$ V. (b)
Spin polarization of the collector current versus time $t$ at
$V = 0.32$ V. The oscillations frequency is 0.25 GHz.}
\label{fig:Delta}
\end{figure}

Figure~\ref{fig:IV} shows the current-voltage $(I-V)$
characteristics of the investigated structure. We obtain the typical
sawtooth-like structure, as can be expected for a multiwell
heterostructure. However, in a broad voltage range from $V = 0.22$ V to 
$V =0.42$ V the current does not reach a steady state value; instead
self-sustained current oscillations occur. In this region also the spin polarization of
the collector current $J_{N\rightarrow N+1}$ is oscillating between
positive and negative values realizing in this way a spin oscillator
device. The effect could serve to generate ac spin currents 
for investigating ac spin injection phenomena.
The maximum amplitude is found at $V=0.32$ V, as illustrated
in Fig.~\ref{fig:Pj}. Along with the current, also the well
magnetization is oscillating, as shown in Fig.~\ref{fig:Delta}.
The phase shift between the magnetizations (one well points into up-direction while the other is
down) is almost independent of the applied bias.

%
%

%
%
By analyzing the local spin densities and the flowing tunneling
currents in-between adjacent wells a qualitative understanding for
the underlying mechanism of these oscillations can be found: (i) In
the bias region, in which self-sustained oscillations are occurring,
the currents $J_{1\rightarrow 2}$ and $J_{2\rightarrow3}$ are
governed by tunneling from the ground state to the first excited
subband of the neighboring wells, as shown schematically in
Fig.~\ref{fig:scheme}. (ii) According to an initial small positive
energy splitting, spin up particles accumulate in the second well
and, hence, $\Delta_2$ increases. (iii) During this accumulation,
particles start to tunnel to the third well. At some point the spin
extraction to the third well overwhelms the spin injection from the
first well. Therefore, with time the spin up particles in the second
well are completely extracted and spin down particles become the
majority species therein. This spin inversion comes along with an
inversion of the exchange splitting $\Delta_2\propto
(n_\uparrow-n_\downarrow)$ from positive to negative values. (iv)
Now the whole process starts from the beginning with the role of
spin up and spin down particles being interchanged.

The frequency of the oscillations is of the order of about 0.25 GHz and
depends strongly, as can be expected, on
the particle tunneling time $\tau_t$, the spin relaxation time
$\tau_s$, and on the response time of the Mn-ions $\tau_\Delta$.
Since $\tau_\Delta$ is usually the largest time-scale in the
problem, it determines the final frequency of the oscillations.

In summary, we have shown that in ferromagnetic multiwell structures
an interesting dynamic interplay of resonant tunneling and carrier
mediated ferromagnetism emerges, which results in stable
self-sustained current oscillations. The underlying mechanism based
on spin inversion is completely different from the one in paramagnetic
multiwell structures, in which moving charge dipoles are formed. The
ferromagnetic heterostructure may be useful as a spin oscillator
device.

This work has been supported by the DFG SFB 689 and the FWF project
P21289-N16.



\end{document}